\documentclass[twocolumn,tighten]{aastex631}
\usepackage{color}
\newcommand{\red}{\textcolor{black}}

\usepackage{multirow}

\received{August 5, 2023}
\revised{September 17, 2023}
\accepted{October 2, 2023}

\shortauthors{Sano, Yamane, van Loon et al. (2023)}
\usepackage{comment}
\defcitealias{2018ApJ...863...55Y}{Paper~I}

\begin{document}
\title{ALMA Observations of Supernova Remnant N49 in the Large Magellanic Cloud.\\II. Non-LTE Analysis of Shock-heated Molecular Clouds}

\author[0000-0003-2062-5692]{H. Sano}
\affiliation{Faculty of Engineering, Gifu University, 1-1 Yanagido, Gifu 501-1193, Japan: sano.hidetoshi.w4@f.gifu-u.ac.jp}
\affiliation{Center for Space Research and Utilization Promotion (c-SRUP), Gifu University, 1-1 Yanagido, Gifu 501-1193, Japan}

\author[0000-0001-8296-7482]{Y. Yamane}
\affiliation{Department of Physics, Nagoya University, Furo-cho, Chikusa-ku, Nagoya 464-8601, Japan: yamane.y@a.phys.nagoya-u.ac.jp}

\author[0000-0002-1272-3017]{J. Th. van Loon}
\affiliation{Lennard-Jones Laboratories, Keele University, Staffordshire ST5 5BG, UK: j.t.van.loon@keele.ac.uk}

\author[0000-0002-2026-8157]{K. Furuya}
\affiliation{National Astronomical Observatory of Japan, Mitaka, Tokyo 181-8588, Japan: kenji.furuya@nao.ac.jp}

\author[0000-0002-8966-9856]{Y. Fukui}
\affiliation{Department of Physics, Nagoya University, Furo-cho, Chikusa-ku, Nagoya 464-8601, Japan: yamane.y@a.phys.nagoya-u.ac.jp}

\author[0000-0001-5609-7372]{R. Z. E. Alsaberi}
\affiliation{Western Sydney University, Locked Bag 1797, Penrith South DC, NSW 1797, Australia}

\author[0000-0003-0890-4920]{A. Bamba}
\affiliation{Department of Physics, Graduate School of Science, The University of Tokyo, 7-3-1 Hongo, Bunkyo-ku, Tokyo 113-0033, Japan}
\affiliation{Research Center for the Early Universe, School of Science, The University of Tokyo, 7-3-1 Hongo, Bunkyo-ku, Tokyo 113-0033, Japan}
\affiliation{Trans-Scale Quantum Science Institute, The University of Tokyo, Tokyo  113-0033, Japan}

\author[0000-0003-2735-3239]{R. Enokiya}
\affiliation{National Astronomical Observatory of Japan, Mitaka, Tokyo 181-8588, Japan: kenji.furuya@nao.ac.jp}
\affiliation{Faculty of Engineering, Gifu University, 1-1 Yanagido, Gifu 501-1193, Japan: sano.hidetoshi.w4@f.gifu-u.ac.jp}

\author[0000-0002-4990-9288]{M. D. Filipovi{\'c}}
\affiliation{Western Sydney University, Locked Bag 1797, Penrith South DC, NSW 1797, Australia}

\author[0000-0002-4663-6827]{R. Indebetouw}
\affiliation{Department of Astronomy, University of Virginia, Charlottesville, VA 22904, USA}
\affiliation{National Radio Astronomy Observatory, 520 Edgemont Road, Charlottesville, VA 22903, USA}

\author[0000-0003-2208-7584]{T. Inoue}
\affiliation{Department of Physics, Konan University, 8-9-1 Okamoto, Higashinada, Kobe, Hyogo 658–8501, Japan}

\author[0000-0001-7813-0380]{A. Kawamura}
\affiliation{National Astronomical Observatory of Japan, Mitaka, Tokyo 181-8588, Japan: kenji.furuya@nao.ac.jp}

\author[0000-0002-8231-0963]{M. Laki\'{c}evi\'{c}}
\affiliation{Astronomical Observatory Belgrade; Volgina 7, 11060 Belgrade, Serbia}

\author[0000-0003-1413-1776]{C. J. Law}
\altaffiliation{NASA Hubble Fellowship Program Sagan Fellow}
\affiliation{Department of Astronomy, University of Virginia, Charlottesville, VA 22904, USA}
\affiliation{Center for Astrophysics $|$ Harvard \& Smithsonian, 60 Garden St., Cambridge, MA 02138, USA}

\author{N. Mizuno}
\affiliation{National Astronomical Observatory of Japan, Mitaka, Tokyo 181-8588, Japan}

\author[0000-0002-9552-3570]{T. Murase}
\affiliation{Faculty of Engineering, Gifu University, 1-1 Yanagido, Gifu 501-1193, Japan: sano.hidetoshi.w4@f.gifu-u.ac.jp}

\author[0000-0001-7826-3837]{T. Onishi}
\affiliation{Department of Physics, Graduate School of Science, Osaka Metropolitan University, 1-1 Gakuen-cho, Naka-ku, Sakai, Osaka 599-8531, Japan}

\author[0000-0003-3900-7739]{S. Park}
\affiliation{Department of Physics, University of Texas at Arlington, Box 19059, Arlington, TX 76019, USA}

\author[0000-0003-1415-5823]{P. P. Plucinsky}
\affiliation{Center for Astrophysics $|$ Harvard \& Smithsonian, 60 Garden St., Cambridge, MA 02138, USA}

\author[0000-0003-3643-839X]{J. Rho}
\affiliation{SETI Institute, 189 N. Bernardo Avenue, Suite 200, Mountain View, CA 94043, USA}

\author[0000-0002-3880-2450]{A. M. S. Richards}
\affiliation{JBCA, Department of Physics and Astronomy, University of Manchester, UK}

\author[0000-0002-9516-1581]{G. Rowell}
\affiliation{School of Physical Sciences, The University of Adelaide, North Terrace, Adelaide, SA 5005, Australia}

\author[0000-0001-5302-1866]{M. Sasaki}
\affiliation{Dr. Karl Remeis-Sternwarte, Erlangen Centre for Astroparticle Physics, Friedrich-Alexander-Universit\"{a}t Erlangen-N\"{u}rnberg, Sternwartstra$\beta$e 7, D-96049 Bamberg, Germany}

\author[0000-0002-0070-3246]{J. Seok}
\affiliation{Department of Physics and Astronomy, University of Missouri, Columbia, MO 65211, USA}
\affiliation{Key Laboratory of Optical Astronomy, National Astronomical Observatories, Chinese Academy of Sciences, Beijing 100012, People’s Republic of China}

\author[0000-0003-3347-7094]{P. Sharda}
\affiliation{Leiden Observatory, Leiden University, P.O. Box 9513, NL-2300 RA Leiden, The Netherlands}

\author[0000-0002-8057-0294]{L. Staveley-Smith}
\affiliation{International Centre for Radio Astronomy Research (ICRAR), University of Western Australia, 35 Stirling Highway, Crawley, WA 6009, Australia}
\affiliation{ARC Centre of Excellence for All Sky Astrophysics in 3 Dimensions (ASTRO 3D), Australia}

\author[0000-0002-8152-6172]{H. Suzuki}
\affiliation{Department of Physics, Konan University, 8-9-1 Okamoto, Higashinada, Kobe, Hyogo 658–8501, Japan}

\author[0000-0001-7380-3144]{T. Temim}
\affiliation{Princeton University, 4 Ivy Ln, Princeton, NJ 08544, USA}

\author[0000-0002-2062-1600]{K. Tokuda}
\affiliation{National Astronomical Observatory of Japan, Mitaka, Tokyo 181-8588, Japan}
\affiliation{Department of Earth and Planetary Sciences, Faculty of Science, Kyushu University, Nishi-ku, Fukuoka 819-0395, Japan}

\author[0000-0002-2794-4840]{K. Tsuge}
\affiliation{Department of Physics, Graduate School of Science, The University of Tokyo, 7-3-1 Hongo, Bunkyo-ku, Tokyo 113-0033, Japan}

\author[0000-0002-1411-5410]{K. Tachihara}
\affiliation{Department of Physics, Nagoya University, Furo-cho, Chikusa-ku, Nagoya 464-8601, Japan: yamane.y@a.phys.nagoya-u.ac.jp}

\begin{abstract}
We present the first compelling evidence of shock-heated molecular clouds associated with the supernova remnant (SNR) N49 in the Large Magellanic Cloud (LMC). Using $^{12}$CO($J$~=~2--1, 3--2) and $^{13}$CO($J$~=~2--1) line emission data taken with the Atacama Large Millimeter/Submillimeter Array, we derived the H$_2$ number density and kinetic temperature of eight $^{13}$CO-detected clouds using the large velocity gradient approximation at a resolution of $3\farcs5$ ($\sim$0.8~pc at the LMC distance). The physical properties of the clouds are divided into two categories: three of them near the shock front show the highest temperatures of $\sim$50~K with densities of $\sim$500--700~cm$^{-3}$, while other clouds slightly distant from the SNR have moderate temperatures of $\sim$20~K with densities of $\sim$800--1300~cm$^{-3}$. The former clouds were heated by supernova shocks, but the latter were dominantly affected by the cosmic-ray heating. These findings are consistent with the efficient production of X-ray recombining plasma in N49 due to thermal conduction between the cold clouds and hot plasma. We also find that the gas pressure is roughly constant except for the three shock-engulfed clouds inside or on the SNR shell, suggesting that almost no clouds have evaporated within the short SNR age of $\sim$4800~yr. This result is compatible with the shock-interaction model with dense and clumpy clouds inside a low-density wind bubble.
\end{abstract}


\section{Introduction}
Supernova remnants (SNRs) have a significant influence on the interstellar medium (ISM) via supernova shocks, injection of heavy elements, and cosmic-ray acceleration. In particular, shock heating and compression of a gaseous medium play an essential role not only in the evolution of the ISM but also in regulating both star formation and the structural evolution of galaxies \citep[e.g.,][]{2015A&A...580A..49I,2016MNRAS.459.3460K}. Several pioneering observations have shown evidence of shock-heated molecular clouds at dozens of kelvins in the vicinity of Galactic SNRs using the non-local thermodynamic equilibrium (non-LTE) approximation \citep[e.g.,][]{1977A&A....54..889C,1987A&A...173..337W}. \cite{1998ApJ...505..286S} presented enhanced $^{12}$CO~$J$~=~2--1/1--0 intensity ratios of $\gtrsim$ 1--3 in shocked molecular clouds in Galactic SNRs W44 and IC 443, and derived typical kinetic temperatures of $\sim$40--80~K in these clouds. Subsequent molecular-line observations also established the presence of supernova-shocked warm clouds using non-LTE codes, and discussed their relation to star formation and/or high-energy phenomena \citep[e.g.,][]{1999ApJ...511..836R,2014A&A...569A..81A,2015ApJ...812...44R,2017ApJ...834...12R,2019ApJ...887...79H,2021ApJ...919..123S,2021ApJ...923...15S,2022A&A...668A.180M}. More recently, an unusually high HCO$^+$/CO ratio has also been observed in molecular clouds interacting with SNRs, possibly due to high cosmic-ray ionization rates \citep{2022ApJ...931..144Z}. On the other hand, there have been no such studies on extragalactic SNRs especially in the Large Magellanic Cloud (LMC), despite the advantages of very little contamination along the line of sight \citep[a small inclination angle of the LMC disk $\sim$20$^{\circ}$--30$^{\circ}$, e.g.,][]{2010A&A...520A..24S} and its known distance \citep[$\sim$$50 \pm 1.3$~kpc,][]{2013Natur.495...76P}.

LHA~120-N49 (hereafter N49) in the LMC is one of the best extragalactic SNRs to study shock-heating and compression of a gaseous medium because it is believed to be associated with molecular clouds \citep[e.g.,][]{1997ApJ...480..607B,2010AJ....139...68V,2015ApJ...808...77U}. The bright radio shell of N49 with centrally filled thermal X-rays is a typical characteristic of a mixed-morphology SNR \citep{1998ApJ...503L.167R}. Although N49 has an apparent diameter of only $\sim$$75''$ \citep[or $\sim$18.2~pc, see][]{2017ApJS..230....2B}, Chandra resolved its filamentary diffuse emission and bright knots of X-rays at a resolution of $\sim$0\farcs5 \citep[e.g.,][]{2003ApJ...586..210P,2012ApJ...748..117P}. The Sedov age of N49 was estimated to $\sim$4800~yr \citep{2012ApJ...748..117P}. The nearby star-forming activity and spectral characteristics of the SNR generally support a massive progenitor star for N49 \citep[][]{2004ApJ...609L..13K,2009ApJ...700..727B,2014ApJ...785L..27Y}. The projection of the magnetar candidate SGR~0526$-$66 within the boundary of N49 may support a core-collapse origin for N49, although the physical association between N49 and SGR~0526$-$66 is not conclusive \citep[e.g.,][]{2001ApJ...559..963G,2012ApJ...748..117P}.  Thus, while the Type Ia origin may not be completely ruled out based on the Si/S ejecta abundance ratio \citep[e.g.,][]{2012ApJ...748..117P}, a core-collapse origin from the explosion of a massive progenitor is generally considered for N49.

A giant molecular cloud (GMC) in the N49 region was first discovered by \cite{1997ApJ...480..607B} using the $^{12}$CO($J$~=~2--1) emission line with the Swedish-ESO Submillimeter Telescope at a resolution of $\sim$$23''$ or $\sim$5.5~pc. The GMC has a diameter of $\sim$14~pc and lies on the southeastern shell of N49, in the direction in which the SNR is brightest in both the radio continuum and X-rays. The authors concluded that the GMC is possibly associated with the SNR. Subsequent Mopra observations using the $^{12}$CO($J$~=~1--0) emission line also supported this idea \citep[][]{2017AIPC.1792d0038S}. An important breakthrough came from observations of the $^{12}$CO($J$~=~1--0) emission line by \cite{2018ApJ...863...55Y} \citepalias[hereafter][]{2018ApJ...863...55Y} using the Atacama Large Millimeter/Submillimeter Array (ALMA). Thanks to the high angular resolution of ALMA of $3\farcs2 \times 2\farcs3$ (or 0.78~pc $\times$ 0.56~pc), the GMC was resolved into 21 molecular clouds. At least three of them are limb-brightened on sub-parsec scales in both hard X-rays and radio continuum, suggesting shock-cloud interactions \citep[e.g.,][]{2009ApJ...695..825I,2012ApJ...744...71I,2010ApJ...724...59S}.

In this paper, we derive the H$_2$ number density and kinetic temperature of molecular clouds in the vicinity of the SNR N49 using ALMA observations of $^{12}$CO($J$~=~1--0, 2--1, 3--2) and $^{13}$CO($J$~=~2--1) and a non-LTE analysis. Our findings for the shock- and/or cosmic-ray-heated clouds provide a new perspective on the ISM surrounding extragalactic SNRs for the first time. In Section~\ref{observations}, we present the observational details and data reduction of the ALMA CO data as well as archival X-ray data. Section~\ref{results} gives an overview of CO distributions and their physical properties. In Sections~\ref{discussion} and \ref{summary} we discuss and summarize our findings.

\section{Observations and Data Reduction}\label{observations}
\subsection{ALMA CO}
We carried out observations of $^{12}$CO($J$~=~3--2) emission line using the ALMA Atacama Compact Array Band~7 in Cycle 8 (PI: Hidetoshi Sano, \#2021.2.00008.S). We used 10--11 antennas of the 7~m array in 2022 August 22, 27, 30, and September 2. We also used 3--4 antennas of the Total Power (TP) array in 2022 August 2, 4, 7, 8, 13, 17, 18, 20, and 27. The effective observed area was an approximately $1\farcm2 \times 1\farcm3$ rectangular region centered at ($\alpha_\mathrm{J2000}$, $\delta_\mathrm{J2000}) = (05^\mathrm{h}26^\mathrm{m}04\fs82$, $-66\arcdeg05\arcmin23\fs8$). The baseline range of the 7~m array was from 8.9 to 91.0~m, corresponding to $u$--$v$ distances from 10.2 to 105.0 $k\lambda$ at 345.796~GHz. Bandpass and flux calibration were carried out using quasar J0538$-$4405 and we used quasar J0601$-$7036 as a phase calibrator.

Observations of $^{12}$CO($J$~=~2--1) and $^{13}$CO($J$~=~2--1) emission lines were carried out using ALMA Band~6 in Cycle 3 (PI: Jacco Th. van Loon, \#2015.1.01195.S). We used 45--46 antennas of the 12~m array in 2016 January 27 and 28; 8--10 antennas of the 7~m array in 2016 August 30, 31, September 2, and 4. The effective observed area was an approximately $1' \times 1'$ rectangular region centered at ($\alpha_\mathrm{J2000}$, $\delta_\mathrm{J2000}) = (05^\mathrm{h}26^\mathrm{m}04\fs00$, $-66\arcdeg05\arcmin20\fs0$). The combined baseline range of the 12 and 7~m arrays was from 8.9 to 331.0~m, corresponding to $u$--$v$ distances from 6.8 to 254.5 $k\lambda$ at 230.538~GHz. The bandpass calibration was carried out by observing three quasars J0519$-$4546, J0522$-$3627, and J0006$-$0623. Another two quasars, J0529$-$7245 and J0601$-$7036, were observed as phase calibrators. We also used Uranus, J0538$-$4405, and J0601$-$7036 as flux calibrators. 

Data reduction and imaging were performed using the Common Astronomy Software Applications package \citep[CASA, version 5.4.1;][]{2022PASP..134k4501C}. We applied the multiscale CLEAN algorithm with the natural weighting scheme \citep[][]{2008ISTSP...2..793C}. Here we used the \texttt{tclean} task to make the images. We then combined the cleaned 7~m data and the calibrated TP array data for CO($J$~=~3--2) and the cleaned 12 and 7~m array data for CO($J$~=~2--1) using the \texttt{feather} task. The final beam size of the feathered image was $3\farcs48 \times 2\farcs76$ with a position angle of 33\fdg7 for the $^{12}$CO($J$~=~3--2) emission line; $1\farcs73 \times 1\farcs10$ with a position angle of 76\fdg7 for the $^{12}$CO($J$~=~2--1) emission line; and $1\farcs82 \times 1\farcs14$ with a position angle of 77\fdg0 for the $^{13}$CO($J$~=~2--1) emission line. The typical RMS noise was $\sim$0.11~K at a velocity resolution of 0.4~km~s$^{-1}$ for the $^{12}$CO($J$~=~3--2) emission line, and $\sim$0.09~K at a velocity resolution of 0.4~km~s$^{-1}$ for both the $^{12}$CO($J$~=~2--1) and $^{13}$CO($J$~=~2--1) emission lines. The primary beam correction was applied with a threshold of 0.2 for $^{12}$CO($J$~=~2--1) and 0.5 for $^{12}$CO($J$~=~3--2) and $^{13}$CO($J$~=~2--1).

In order to take an intensity ratio of two different CO $J$-transitions, we also used the primary-beam-corrected, $^{12}$CO($J$~=~1--0) cleaned data published in \citetalias[][]{2018ApJ...863...55Y}. The beam size was $3\farcs19 \times 2\farcs25$ with a position angle of 66\fdg14. The typical RMS noise was $\sim$0.07~K at a velocity resolution of 0.4~km~s$^{-1}$.

\subsection{Chandra X-Rays}
We utilized archival X-ray data obtained by Chandra with the Advanced CCD Imaging Spectrometer S-array (ACIS-S3). The observation IDs were 747, 1041, 1957, 2515, 10123, 10806, 10807, and 10808, and were published in previous papers \citep[][]{2003ApJ...586..210P,2012ApJ...748..117P,2020ApJ...894...17P,2018ApJ...863...55Y,2019A&A...629A..51Z}. We utilized Chandra Interactive Analysis of Observations software \citep[CIAO version 4.12;][]{2006SPIE.6270E..1VF} with CALDB 4.9.1 \citep[][]{2007ChNew..14...33G} for the data reduction. All of the data were reprocessed using the task {\texttt{chandra\_repro}}. The total effective exposure reached $\sim$197~ks for the eastern rim of the SNR, where the shock-cloud interactions occur. We then made an exposure-corrected, energy-ﬁltered image using the task {\texttt{ﬂuximage}} in the energy band of 0.5--7.0~keV as a broad-band X-ray image. 

\begin{figure*}[]
\begin{center}
\includegraphics[width=\linewidth,clip]{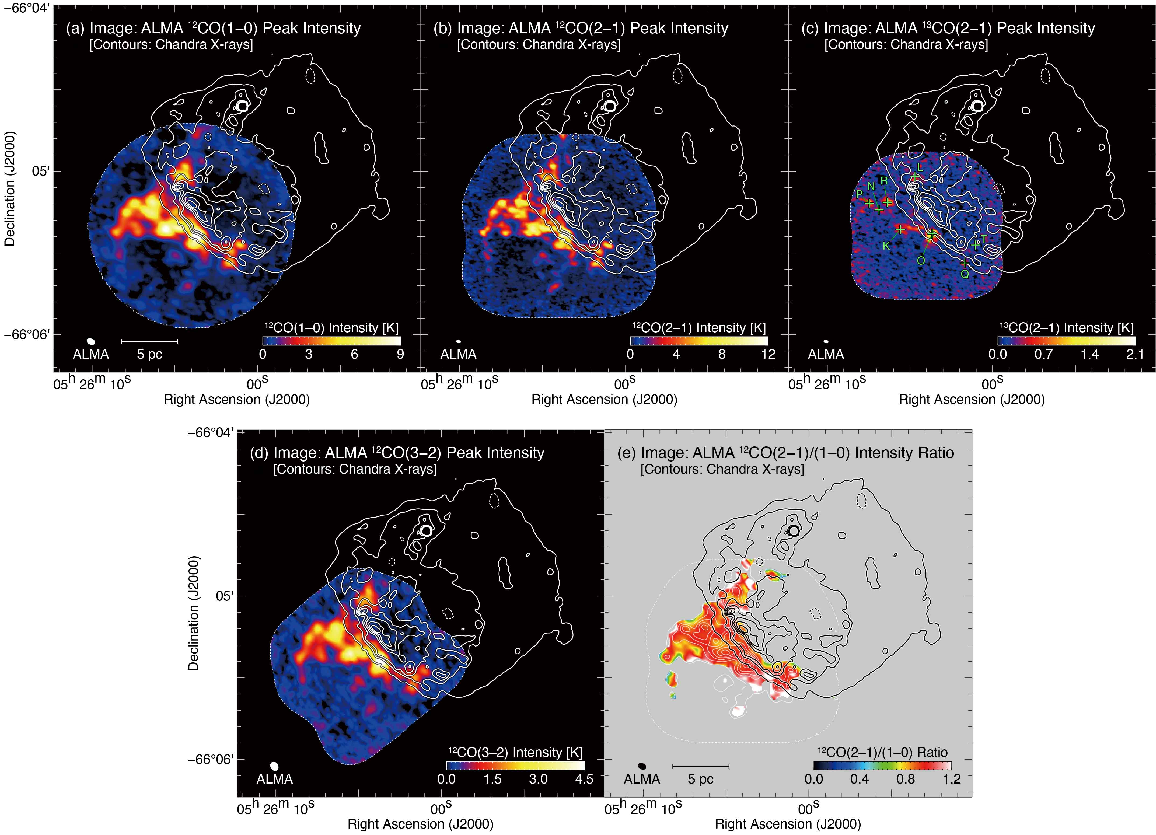}
\caption{Peak intensity maps of (a) $^{12}$CO($J$~=~1--0), (b) $^{12}$CO($J$~=~2--1), (c) $^{13}$CO($J$~=~2--1), and (d) $^{12}$CO($J$~=~3--2). (e) Intensity ratio map of $^{12}$CO($J$~=~2--1)/$^{12}$CO($J$~=~1--0) (hereafter $R_\mathrm{2-1/1-0}$). The velocity range of CO is from 285.5 to 287.8 km s$^{-1}$. Superposed contours represent Chandra X-ray intensity in the energy band of 0.5--7.0~keV. The lowest contour level and the contour intervals are $5 \times 10^{-7}$ and $3 \times 10^{-6}$ photons pixel$^{-1}$ s$^{-1}$, respectively. The regions enclosed by dashed lines indicate the observed areas using ALMA. The CO peaks K, L, N, O--R, and T defined by \citetalias[][]{2018ApJ...863...55Y} are also indicated in (c). When we take the intensity ratio of $R_\mathrm{2-1/1-0}$, the $^{12}$CO($J$~=~2--1) data were smoothed with a two-dimensional Gaussian kernel to match the effective beam size of the $^{12}$CO($J$~=~1--0) data. The gray shaded area in (e) represents the $^{12}$CO($J$~=~1--0) and/or $^{12}$CO($J$~=~2--1) data showing significance lower than $\sim$10$\sigma$.}
\label{fig1}
\end{center}
\end{figure*}%

\section{Results}\label{results}
\subsection{Distributions of CO and X-Rays}
Figures~\ref{fig1}a--\ref{fig1}d show the peak intensity maps of $^{12}$CO($J$~=~1--0), $^{12}$CO($J$~=~2--1), $^{13}$CO($J$~=~2--1), and $^{12}$CO($J$~=~3--2) at $V_\mathrm{LSR} = 285.5$--287.8~km~s$^{-1}$, corresponding to the shocked components in N49 (\citetalias[][]{2018ApJ...863...55Y}). The clouds detected in $^{12}$CO($J$~=~2--1) and $^{12}$CO($J$~=~3--2) are situated along the southeastern edge of the X-ray shell and their spatial distribution is consistent with that seen in  $^{12}$CO($J$~=~1--0) in \citetalias[][]{2018ApJ...863...55Y}. 

In comparison to $^{12}$CO, the $^{13}$CO clouds are distributed more sparsely (see Figure~\ref{fig1}c). 
Generally, in the LMC, both pre-star formation and active star-forming clouds exhibit relatively similar distributions for $^{12}$CO and $^{13}$CO \citep[e.g.,][]{2019ApJ...885...50W}, often forming prominent filamentary structures \citep[e.g.,][]{2019ApJ...886...14F,2019ApJ...886...15T}. The compact, peculiar nature of the $^{13}$CO distribution in N49 is distinct from the above-mentioned interstellar trends, and unique to this SNR. Eight of them, K, L, N, O--R, and T, are detected in $^{13}$CO($J$~=~2--1) with $3\sigma$ or higher significance. The fundamental physical properties of $^{13}$CO-detected clouds (position, brightness temperature, center velocity, spectral linewidth, cloud size, virial mass) are summarized in Table~\ref{tab1}. In the present paper, we focus on these clouds to estimate their kinetic temperature and H$_2$ number density.

\begin{deluxetable*}{lcccccccccccc}
\tablewidth{\linewidth}
\tabletypesize{\scriptsize}
\tablecaption{Physical Properties of $^{13}$CO Detected Clouds in SNR N49}
\label{tab1}
\tablehead{\\
\multicolumn{1}{c}{\multirow{2}{*}{Name}} & \multirow{2}{*}{$\alpha_\mathrm{J2000}$} & \multirow{2}{*}{$\delta_\mathrm{J2000}$} & \multicolumn{3}{c}{$T_\mathrm{b}$} & \multirow{2}{*}{$V_{\mathrm{peak}}$} &  \multirow{2}{*}{$\bigtriangleup V$ }& \multicolumn{1}{c}{\multirow{2}{*}{Size}}  & \multicolumn{1}{c}{\multirow{2}{*}{$M_\mathrm{vir}$}} & \multicolumn{1}{c}{\multirow{2}{*}{$R_\mathrm{dist}$}} & \multirow{2}{*}{$T_\mathrm{kin}$} & \multirow{2}{*}{$n$(H$_2$)}\\[2pt]
\cline{4-6}\\[-8pt]
&&& $^{12}$CO(3--2) & $^{12}$CO(2--1) & $^{13}$CO(2--1) &&&&&& &\\[1pt]
 & (h m s) & ($^{\circ}$ $'$ $''$) & (K) & (K) & (K) & (km $\mathrm{s^{-1}}$) & (km $\mathrm{s^{-1}}$)  & \multicolumn{1}{c}{(pc)} & \multicolumn{1}{c}{($M_\sun $)} & \multicolumn{1}{c}{(pc)} & (K) & ($10^3$~cm$^{-3}$)\\
\multicolumn{1}{c}{(1)} & (2) & (3) & (4) & (5) & (6) & \multicolumn{1}{c}{(7)} & \multicolumn{1}{c}{(8)} & \multicolumn{1}{c}{(9)} & (10) & (11) & (12) & (13)
}
\startdata
K ...... & 5$:$26$:$5.5 & $-$66$:$05$:$21 & $3.60$ & $6.59$ & $0.35$ & $285.5$ & $0.83$ & 1.3 & \phantom{0}80 & 10.7 & $45^{+14}_{-9\phantom{0}}$ & $0.53^{+0.05}_{-0.06}$\\
L ...... & 5$:$26$:$4.7 & $-$66$:$05$:$02 & $2.00$ & $3.75$ & $0.18$ & $286.3$ & $1.72$ & 1.2 & 340 & \phantom{0}7.7 & $42^{+26}_{-13}$ & $0.72^{+0.13}_{-0.14}$\\
N ...... & 5$:$26$:$6.7 & $-$66$:$05$:$15 & $2.58$ & $4.67$ & $0.40$ & $286.4$ & $0.40$ & 0.8 & \phantom{0}10 & 11.6 & $21^{+4\phantom{0}}_{-3\phantom{0}}$ & $0.78^{+0.02}_{-0.04}$\\
O ...... & 5$:$26$:$3.7 & $-$66$:$05$:$23 & $4.91$ & $8.26$ & $0.91$ & $286.2$ & $0.69$ & 1.2 & \phantom{0}50 & \phantom{0}8.8 & $19^{+1\phantom{0}}_{-2\phantom{0}}$ & $1.10^{+0.01}_{-0.01}$\\
P ...... & 5$:$26$:$7.5 & $-$66$:$05$:$12 & $2.17$ & $3.70$ & $0.36$ & $286.4$ & $1.00$ & 1.1 & 100 & 12.2 & $19^{+4\phantom{0}}_{-3\phantom{0}}$ & $1.32^{+0.01}_{-0.03}$\\
Q ...... & 5$:$26$:$1.7 & $-$66$:$05$:$34 & $1.82$ & $3.64$ & $0.25$ & $286.5$ & $0.90$ & 1.0 & \phantom{0}80 & \phantom{0}9.6 & $21^{+6\phantom{0}}_{-4\phantom{0}}$ & $0.81^{+0.06}_{-0.05}$\\
R ...... & 5$:$26$:$6.3 & $-$66$:$05$:$11 & $3.17$ & $5.49$ & $0.51$ & $286.4$ & $0.86$ & 1.1 & \phantom{0}80 & 10.7 & $20^{+3\phantom{0}}_{-2\phantom{0}}$ & $1.10^{+0.02}_{-0.03}$\\
T ...... & 5$:$26$:$1.0 & $-$66$:$05$:$27 & $1.55$ & $2.94$ & $0.14$ & $286.4$ & $0.57$ & 0.9 & \phantom{0}30 & \phantom{0}7.7 & $46^{+57}_{-18}$ & $0.48^{+0.09}_{-0.14}$\\
\enddata
\tablecomments{Col. (1): Cloud name defined by \citetalias[][]{2018ApJ...863...55Y}. Cols. (2, 3): Position of the maximum intensity of $^{12}$CO($J$~=~1--0) \citepalias[][]{2018ApJ...863...55Y}. Cols. (4--8): Physical properties of CO emission derived from a single Gaussian fitting. All the CO datasets were smoothed to match the beam size of $3\farcs5 \times 3\farcs5$. Cols. (4--6): Peak brightness temperature of $^{12}$CO($J$~=~3--2), $^{12}$CO($J$~=~2--1), and $^{13}$CO($J$~=~2--1) emision. Col. (7): Central velocity of $^{13}$CO($J$~=~2--1) spectra, $V_{\mathrm{peak}}$. Col. (8): Full-width at half-maximum (FWHM) line width of $^{13}$CO($J$~=~2--1) spectra, $\bigtriangleup V$. Col. (9): $^{13}$CO($J$~=~2--1) derived cloud size defined as ($A$/$\pi$)$^{0.5}$ $\times 2 $, where $A$ is the total cloud surface area surrounded by the half intensity of peak integrated intensity contour. Col. (10): Mass of the cloud derived using the virial theorem and $^{13}$CO($J$~=~2--1) properties, $M_\mathrm{vir}$. Col. (11): Radial distance from the geometric center of the SNR at ($\alpha_\mathrm{J2000}$, $\delta_\mathrm{J2000}) = (5^\mathrm{h}25^\mathrm{m}59\fs57$, $-66\arcdeg04\arcmin56\fs4$). Col. (12): Kinetic temperature, $T_\mathrm{kin}$. Col. (12): Number density of molecular hydrogen, $n$(H$_2$).}
\vspace*{-1cm}
\end{deluxetable*}

\subsection{Intensity Ratio of $^{12}$CO($J$~=~2--1)/$^{12}$CO($J$~=~1--0)}
Figure~\ref{fig1}e shows the intensity ratio distribution of $^{12}$CO($J$~=~2--1)/$^{12}$CO($J$~=~1--0) (hereafter $R_\mathrm{2-1/1-0}$). The intensity ratio is a good indicator of the degree of the CO rotational excitation because the $J$~=~2 level is at an excitation energy of $E/k = 16.5$~K, $\sim$10~K above the $J$~=~1 level at $E/k = 5.5$~K \citep[e.g.,][]{1998ApJ...505..286S}. We find a high intensity ratio of $R_\mathrm{2-1/1-0}$ $\sim$1.0--1.2 in the entire molecular cloud associated with the SNR. We also confirmed that the intensity ratio distribution of $^{12}$CO($J$~=~3--2)/$^{12}$CO($J$~=~1--0) is similar to that of $R_\mathrm{2-1/1-0}$. It is noteworthy that no extra heating sources such as OB stars and/or young stellar objects are associated with the molecular clouds in the vicinity of the SNR.

\begin{figure*}[]
\begin{center}
\includegraphics[width=\linewidth,clip]{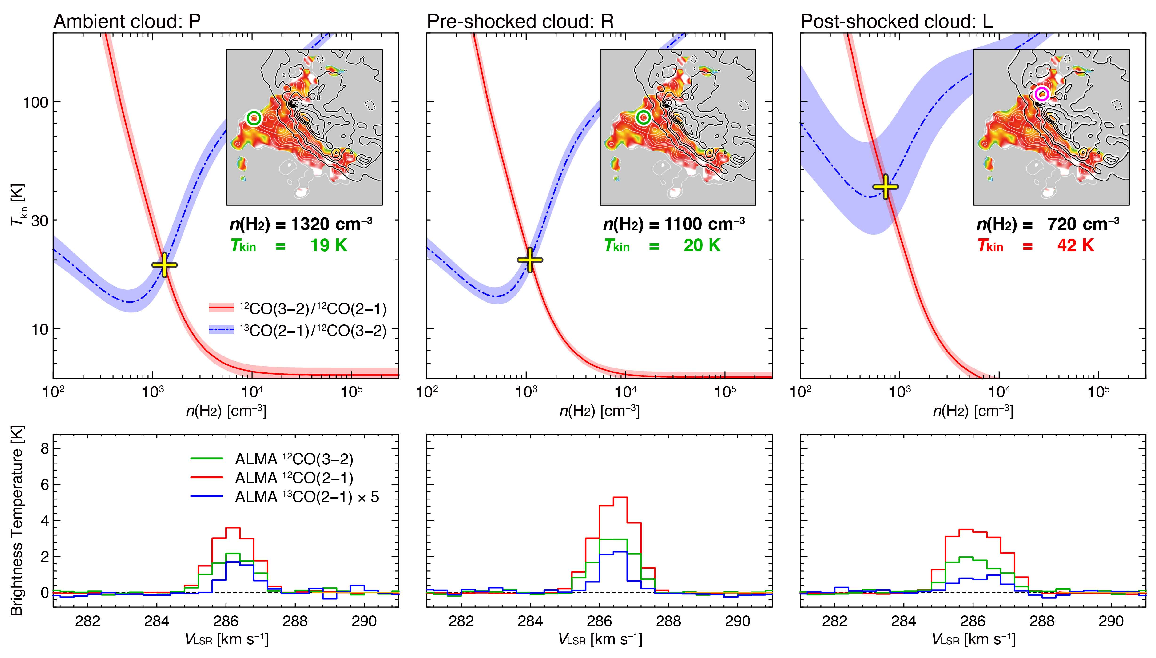}
\caption{Top panels: large velocity gradient results on the number density of molecular hydrogen, $n$(H$_2$), and the kinetic temperature, $T_\mathrm{kin}$, for the ambient cloud P, pre-shocked cloud R, and the post-shocked cloud L. The red lines and blue dashed-dotted lines indicate the intensity ratios of $^{12}$CO($J$~=~3--2)/$^{12}$CO($J$~=2--1) and $^{13}$CO($J$~=~2--1)/$^{12}$CO($J$~=3--2), respectively. The shaded areas surrounding the red and blue lines indicate the $1\sigma$ ranges of each intensity ratio. Yellow crosses represent the best-fit values of $n$(H$_2$) and $T_\mathrm{kin}$ for each cloud. The spatial positions and best-fit values for each cloud are shown in the top-right corners for each panel. Bottom panels: CO intensity profiles for clouds P, R, and L. The physical properties of each cloud are summarized in Table 1.}
\label{fig2}
\end{center}
\vspace*{-1cm}
\end{figure*}%

\begin{figure}[]
\begin{center}
\includegraphics[width=\linewidth,clip]{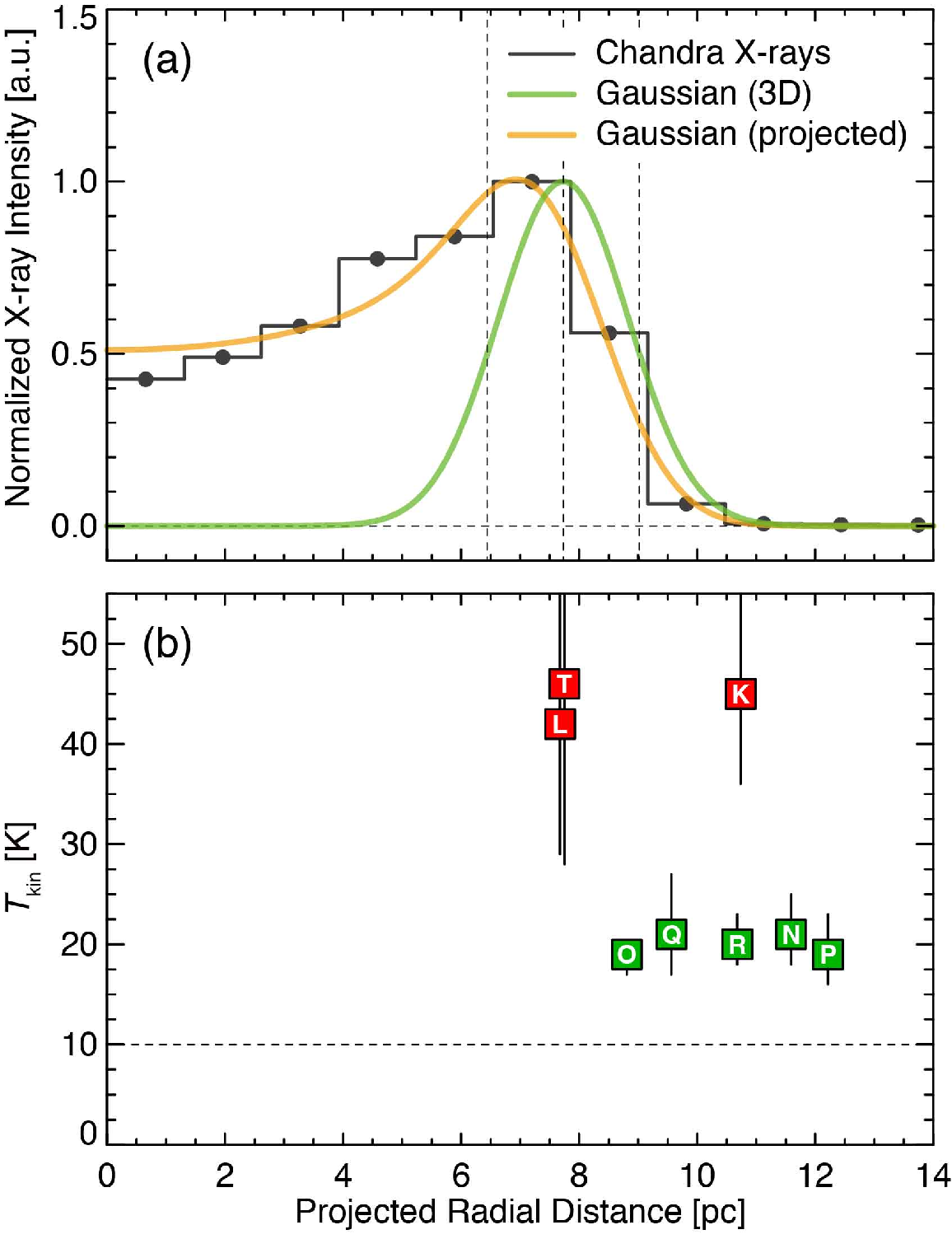}
\caption{(a) Radial profile of the normalized X-ray intensity, centered at ($\alpha_\mathrm{J2000}$, $\delta_\mathrm{J2000}) = (05^\mathrm{h}25^\mathrm{m}59\fs57$, $-66\arcdeg04\arcmin56\fs4$) as the geometric center of the SNR. The steps with ﬁlled circles indicate averaged values of normalized X-ray intensity at each annulus. The green line indicates the three-dimensional Gaussian distribution, and the orange line represents its projected distribution given by the least-squares ﬁtting (see the text). (b) Scatter plot of kinetic temperature $T_\mathrm{kin}$ for each cloud as a function of the projected radial distance. The dashed horizontal line represents $T_\mathrm{kin}$ = 10~K, coresponding to the typical temperature of quiescent molecular clouds. The red and green symbols represent high- and moderate temperature clouds, respectively.}
\label{fig3}
\end{center}
\end{figure}%

\subsection{Large Velocity Gradient Analysis}
In order to estimate the kinetic temperature $T_\mathrm{kin}$ and H$_2$ number density $n$(H$_2$) of molecular clouds in N49, we performed a non-LTE analysis using the large velocity gradient (LVG) approximation \citep[e.g.,][]{1974ApJ...189..441G,1974ApJ...187L..67S}. This non-LTE code calculates the radiative transfer of multiple transitions of atomic and/or molecular lines, assuming an isothermal spherical cloud with uniform photon escape probability and velocity gradient of $dv/dr$. Here, $dv$ is the half-maximum half-width (HMHW) linewidth of the CO line profile derived by least-squares fitting with a Gaussian kernel, and $dr$ is the cloud radius. Since the eight molecular clouds have different cloud sizes and linewidths, we calculated individual $dv/dr$ for each cloud. We also used the abundance ratio of [$^{12}$CO/H$_2$]$~=~1.6 \times 10^{-5}$ and the abundance ratio of isotopes [$^{12}$CO/$^{13}$CO]~=~50 \citep[e.g.,][]{1987ApJ...315..621B,2010PASJ...62...51M,2011AJ....141...73M,2014ApJ...796..123F}. The $^{12}$CO($J$~=1--0) data were not included in the LVG analysis because the lowest-$J$ transition may be subject to self-absorption owing to a large optical depth \citep[e.g.,][]{2010PASJ...62...51M}.

\begin{figure}[]
\begin{center}
\includegraphics[width=\linewidth,clip]{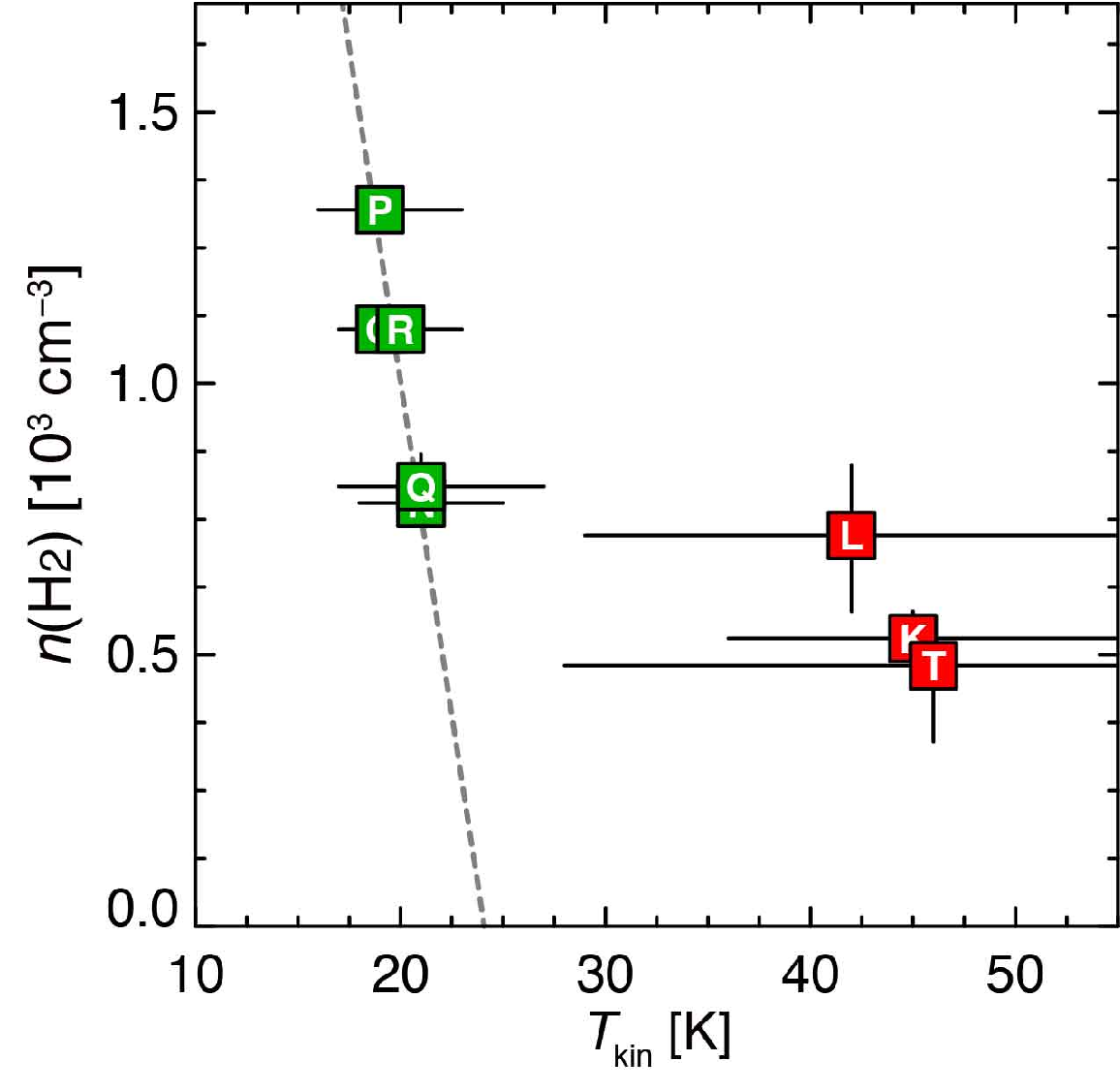}
\caption{Scatter plot between $T_\mathrm{kin}$ and $n$(H$_2$). The dashed line indicates the linear regression using the least-squares method except for the CO clouds L, K, and T (see the text).}
\label{fig4}
\end{center}
\end{figure}%

Figure~\ref{fig2} shows the typical LVG results, $T_\mathrm{kin}$--$n$(H$_2$) relation, and CO spectra for clouds P, R, and L. The post-shocked cloud L on the shock front shows higher kinetic temperature $T_\mathrm{kin}$ $\sim$42~K and relatively lower H$_2$ number density $n$(H$_2$) $\sim$720~cm$^{-3}$, while the ambient cloud P away from the SNR shell shows lower $T_\mathrm{kin}$ $\sim$19~K and moderate $n$(H$_2$) $\sim$1320~cm$^{-3}$. The pre-shocked cloud R lying between them shows similar values of $T_\mathrm{kin}$ and $n$(H$_2$) to the ambient cloud P ($T_\mathrm{kin}$ $\sim$20~K and $n$(H$_2$) $\sim$1100~cm$^{-3}$). This trend is also found in other molecular clouds in N49. It is noteworthy that clouds K, L, and T show lower $^{13}$CO($J$~=~2--1) / $^{12}$CO($J$~=~3--2) ratios than other clouds.

Figure~\ref{fig3} shows the radial distributions of X-ray intensity and $T_\mathrm{kin}$ of the eight molecular clouds from the geometric center of the SNR at ($\alpha_\mathrm{J2000}$, $\delta_\mathrm{J2000}) = (05^\mathrm{h}25^\mathrm{m}59\fs57$, $-66\arcdeg04\arcmin56\fs4$). To calculate the shell diameter and thickness of the X-ray shell in N49, we performed a least-squares fitting using a three-dimensional spherical shell model with a Gaussian function $F(r)$ \citep[e.g.,][]{2017ApJ...843...61S,2021ApJ...923...15S,2022ApJ...938...94A}:
\begin{eqnarray}
F(r) =  A \exp[-(r - r_0\bigr)^2/2\sigma ^2],
\label{eq3}
\end{eqnarray}
where $A$ is the normalization factor of the Gaussian function, $r_0$ is the shell radius, $r$ is the radial distance from the geometric center of the SNR, and $\sigma$ is the standard deviation of the Gaussian function. We obtained a shell radius of $7.7 \pm 1.8$~pc and shell thickness of $2.6 \pm 4.5$~pc as the best-fit parameters of the fitting, where the shell thickness was defined as the FWHM of the Gaussian function.

Figure~\ref{fig4} shows the scatter plot between $T_\mathrm{kin}$ and $n$(H$_2$) for the eight molecular clouds. Clouds N, O, P, Q, and R show a negative correlation which was fitted by the least-squares method using the \texttt{MPFITEXY} routine \citep{2010MNRAS.409.1330W}, while $T_\mathrm{kin}$ of clouds K, L, and T seems to be shifted to higher values compared to the other molecular clouds.

\section{Discussion}\label{discussion}
The observed spatial variation in $T_\mathrm{kin}$ and $n$(H$_2$) of the molecular clouds associated with N49 has many implications for the evolution of the supernova-shocked ISM. First, we argue that the high temperatures of $\sim$40--50~K as seen in clouds K, L, and T are likely caused by shock-heating. According to \citetalias[][]{2018ApJ...863...55Y}, these clouds are tightly interacting with shockwaves and enhance the turbulent magnetic field with bright synchrotron radiation on the surface of the shocked clouds, suggesting that clouds K, L, and T are located inside the SNR or on the shock front. Quantifying this scenario requires a decomposition of the density and velocity distribution of clouds K, L and T which requires even higher-resolution data \citep[e.g.,][]{2021MNRAS.500.1721M,2022MNRAS.509.2180S}. Although clouds O, Q, and R have smaller radial distances than cloud K, this is likely due to projection effects. In addition, the observed cloud temperatures are roughly consistent with those measured in shock-heated molecular clouds associated with several Galactic SNRs \citep[e.g.,][]{1998ApJ...505..286S,2021ApJ...923...15S,2021ApJ...919..123S}. The lack of local excitation sources other than SNR shocks also supports this scenario. If so, this is the first compelling evidence of ``shock-heated'' molecular clouds associated with an extragalactic SNR.

The shock-heating scenario is also consistent with the thermal conduction origin of the recombining plasma (RP) discovered in N49 \citep{2015ApJ...808...77U}. The RP is characterized by a higher ionization temperature compared to the electron temperature, which is not expected in the classical evolution model of thermal plasma in SNRs \citep[see also the review by][]{2020AN....341..150Y}. To achieve the RP state, rapid electron cooling or an increase of the ionization state is needed. One of the possible scenarios is the rapid cooling of electrons via thermal conduction between the shocks and cold/dense clouds \citep[e.g.,][]{2002ApJ...572..897K,2017ApJ...851...73M,2017PASJ...69...30M,2018PASJ...70...35O,2020ApJ...890...62O,2021ApJ...919..123S}. \citetalias[][]{2018ApJ...863...55Y} found that the hard-band X-rays due to radiative recombination are enhanced around the shocked clouds including clouds K, L, and T. The authors concluded that the RP in N49 was efficiently produced by shocked CO clouds due to thermal conduction, which is consistent with the higher kinetic temperatures of clouds K, L, and T. Additional spatially resolved X-ray spectroscopy with high-angular/spectral resolution is needed to confirm this scenario. The Athena mission has the potential to accomplish these \citep[e.g.,][]{2013arXiv1306.2307N}. It is noteworthy that the moderate dust temperature in cold components $\sim$20--30~K (pre-shocked dust) and in warm components $\sim$60~K (post-shocked dust) in N49 is also roughly consistent with our CO results \citep[e.g.,][]{2010A&A...518L.139O,2010AJ....139.1553V,2015ApJ...799...50L}, but the detailed comparison between the dust and CO clouds is beyond the scope of this paper.

Next, we argue that the moderate kinetic temperatures of $\sim$20~K seen in the other clouds N, O, P, Q, and R are likely due to cosmic-ray heating. According to \cite{2009MNRAS.396.1629G}, the diffusion length of cosmic rays $l_\mathrm{diff}$ in units of centimeters can be described as
\begin{eqnarray}
l_\mathrm{diff} = \sqrt{4 D(E, B) t_\mathrm{age}}\;,
\label{eq1}
\end{eqnarray}
where $D(E, B)$ is the diffusion coefficient in units of cm$^{2}$~s$^{-1}$ and $t_\mathrm{age}$ is the SNR age in units of seconds. $D(E, B)$ can be written using the accelerated cosmic-ray energy $E$ and the magnetic field strength $B$ as 
\begin{eqnarray}
D(E, B) = 5 \times 10^{26} \left(\frac{E}{100~\mathrm{MeV}}\right)^{0.5}\left(\frac{B}{10~\mathrm{\mu G}}\right)^{-0.5}\mathrm{}.\;\;\;
\label{eq2}
\end{eqnarray}
Adopting $t_\mathrm{age} = 1.5 \times 10^{11}$~s \citep[$\sim$4800~yr;][]{2012ApJ...748..117P}, $E = 100$~MeV, and $B = 10~\mu$G, we obtain $l_\mathrm{diff} \sim$6~pc which is substantially larger than the maximum separation between cloud P and the shock boundary of the SNR. This indicates that almost all cosmic ray particles accelerated in N49 have reached the entire region of the associated molecular clouds, and hence the cosmic-ray heating works equally well for both the post-shocked and pre-shocked clouds as well as the ambient clouds. In other words, cosmic-ray heating is not sufficient to increase the kinetic temperature of K, L, and T significantly above those of the unshocked clouds (O, Q, R, etc).

\begin{figure}[]
\begin{center}
\includegraphics[width=\linewidth,clip]{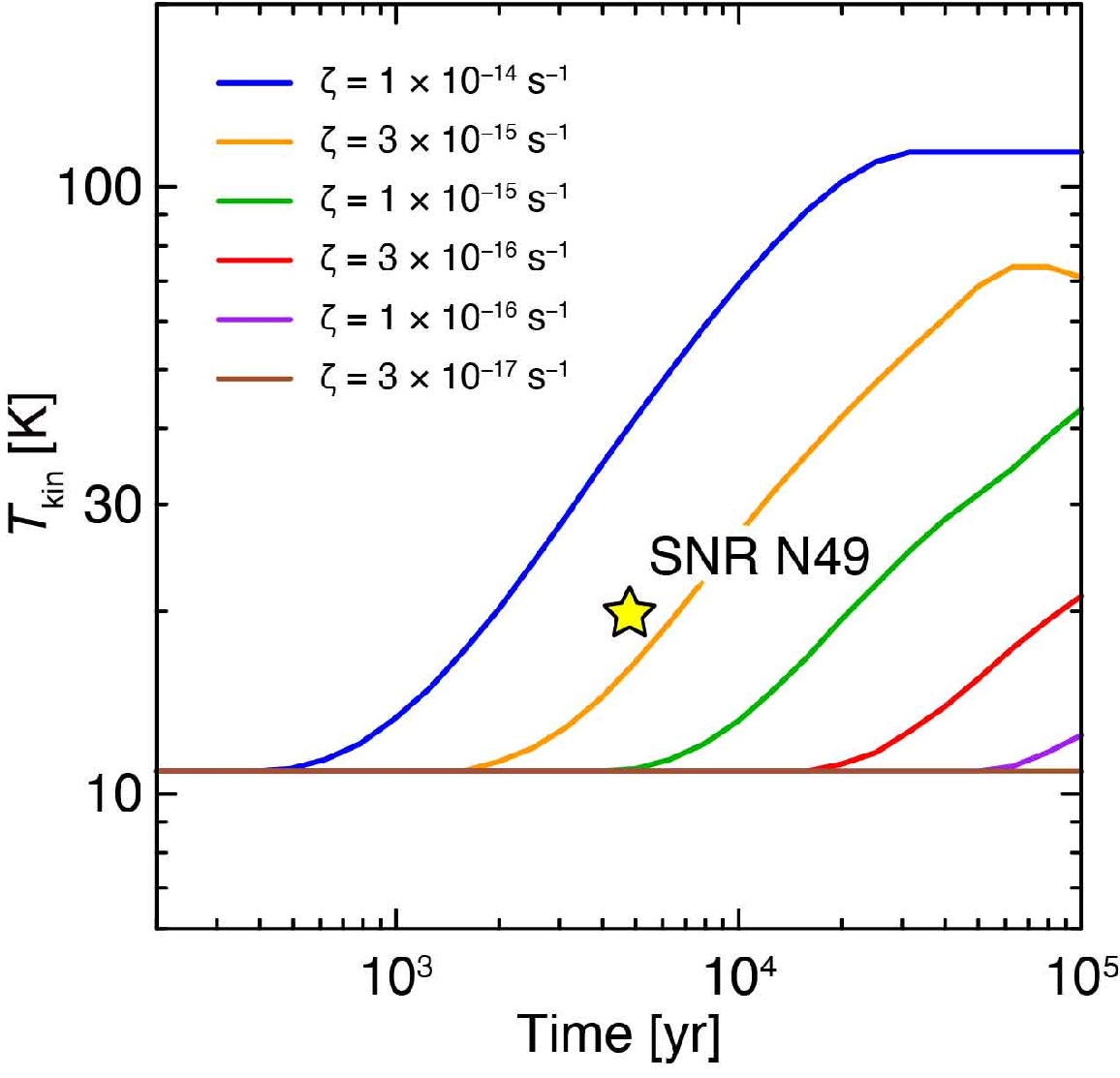}
\caption{Time evolution of kinematic temperature $T_\mathrm{kin}$ at the cloud center after the supernova event that generates SNR N49. Colors indicate the assumed cosmic-ray ionization rate $\zeta$ from $3 \times 10^{-17}$~s$^{-1}$ to $1\times10^{-14}$~s$^{-1}$. The star symbol indicates the observed values of $T_\mathrm{kin}$ for the ambient or pre-shocked clouds at the age of N49 (4800~yr).}
\label{fig5}
\end{center}
\end{figure}%

We also argue that the moderate temperatures of the cosmic-ray-irradiated clouds are naturally explained by the high cosmic-ray ionization rates surrounding the SNR. Figure~\ref{fig5} shows the time evolution of the kinematic temperature $T_\mathrm{kin}$ at the cloud center after the supernova event that generated the LMC SNR N49, which was calculated as a function of cosmic-ray ionization rates $\zeta$ using a photodissociation region (PDR) model \citep[][see the Appendix for further details]{2022ApJ...926..148F}. Since the typical cosmic-ray ionization rates in the vicinity of Galactic SNRs are $\sim$30--300 times higher than the standard Galactic value of $\sim$$10^{-17}$~s$^{-1}$ \citep[e.g.,][]{2010ApJ...724.1357I,2011ApJ...740L...4C,2014A&A...568A..50V}, the observed values of $T_\mathrm{kin}$ $\sim$20~K at pre-shocked and ambient clouds can be explained as the cosmic-ray heating effect. Further follow-up observations using multiple emission lines such as HCO$^{+}$, DCO$^{+}$ \citep[e.g.,][]{2014A&A...568A..50V}, and/or [C{\sc i}] \citep{2023arXiv230615983Y} are needed to estimate the cosmic-ray ionization rate in the vicinity of the LMC SNR N49.

\red{Another interesting point is that these clouds, when exposed to a large number of cosmic rays, likely emit gamma-rays and X-ray line emission. Recent Fermi-LAT analysis of N49 presented a possible detection of hadronic gamma-rays which were produced by proton--proton interactions \citep[][]{2022MNRAS.515.1676C}. The Fe K$\alpha$ line emission at 6.4~keV was also discovered in several Galactic/Magellanic SNRs, which was probably produced by interactions between low-energy cosmic-ray protons and neutral irons in the ISM clouds \citep[e.g.,][]{2018ApJ...854...87N,2019PASJ...71..115N,2009ApJ...691.1854B,2018ApJ...854...71B,2018PASJ...70...35O}. Further gamma-ray and X-ray observations at the high spatial and spectral resolutions using the Cherenkov Telescope Array \citep{2011ExA....32..193A,2019scta.book.....C}, the XRISM mission \citep{2020SPIE11444E..22T}, and the Athena mission \citep[][]{2013arXiv1306.2307N} will allow us to unveil the high-energy radiation processes in N49.} 

Finally, we focus on possible cloud evaporation and/or destruction via shock-cloud interactions. In general, it is often assumed that interstellar gaseous clouds are immediately evaporated by supernova shocks in SNRs \citep[e.g.,][]{2017ApJ...846...77S,2018MNRAS.478.2948W,2019ApJ...875...81Z}. However, this is correct only if the gas density contrast between the cloud and the inter-cloud gas densities is small, typically $\sim$100 or less. The initial ISM conditions before a core-collapse supernova explosion are expected to have a large density fluctuation of $\sim$10$^5$ due to strong stellar winds from a massive progenitor. The low-density inter-cloud medium, such as atomic H{\sc i} gas, was completely swept up by the winds over $\sim$$10^6$~yr and then created a wind bubble with an interior density of $\sim$0.01~cm$^{-3}$ \citep[][]{1977ApJ...218..377W}. On the other hand, molecular clouds can survive wind erosion owing to their high densities, greater than $\sim$$10^3$~cm$^{-3}$. When the shocks encounter the molecular cloud, the shock velocity $V_\mathrm{sh}$ dramatically decelerates to $V_\mathrm{sh} / (n_\mathrm{cloud} / n_0)^{0.5}$, where $n_\mathrm{cloud}$ is the cloud density and $n_0$ is the inter-cloud (ambient) density \citep[see Section 39.3 in][]{2011piim.book.....D}. Adopting the density contrast $n_\mathrm{cloud} / n_0 = 10^5$, $V_\mathrm{sh} = 3000$~km~s$^{-1}$, and a cloud radius of 1~pc, the cloud-crossing time of supernova shocks is $\sim$$10^5$ yr, and hence the cloud can survive the shock destruction within the typical lifetime of a SNR \citep[see also a review by][]{2021Ap&SS.366...58S}.

We argue that the scenario can be applied to the SNR N49. The wind-bubble explosion scenario is consistent with the core-collapse origin of N49 \citep[][]{2001ApJ...559..963G,2004ApJ...609L..13K,2009ApJ...700..727B,2014ApJ...785L..27Y}. The shocked molecular clouds are thought to have survived the erosion because we detect them using CO emission lines. Moreover, the negative correlation between $T_\mathrm{kin}$ and $n$(H$_2$) indicates that clouds N, O, P, Q, and R have constant pressure and thus did not undergo shock evaporation. The enhanced magnetic field via shock-cloud interactions may also suppress thermal conduction and hamper cloud evaporation \citep[e.g.,][]{2008ApJ...678..274O}. Interestingly, clouds K, L, and T located inside or on the edge of the SNR shell are significantly shifted from this pressure-constant line to a higher-pressure region and have slightly reduced number densities $n$(H$_2$). This trend suggests that the clouds completely engulfed by shocks were partially evaporated through shock-cloud interactions. A similar trend is also seen in the shock-engulfed star-forming dense core C embedded within the Galactic SNR RX~J1713.7$-$3946. \cite{2010ApJ...724...59S} presented a slightly steeper density gradient than the typical star-forming cloud core, suggesting that only the surface of the cloud has been stripped out and heated up. In any case, the molecular clouds engulfed by the fast shocks in N49 were shown to decrease in number density rather than increase.

\section{Summary}\label{summary}
We summarize our conclusions as follows.
\begin{enumerate}
\item New $^{12}$CO($J$~=~3--2, 2--1) and $^{13}$CO($J$~=~2--1) observations using ALMA have revealed clumpy distributions of molecular clouds associated with SNR N49 in the LMC at a resolution of $1\farcs73 \times 1\farcs10$ or $\sim$0.3~pc. The clouds completely delineate the southeastern edge of the SNR shell and show a high-intensity ratio of $^{12}$CO($J$~=~2--1)/$^{12}$CO($J$~=~1--0) $\sim$1.0--1.2, suggesting that shock-cloud interactions have occurred.
\item We applied a non-LTE LVG analysis toward the eight molecular clouds that are detected in $^{13}$CO($J$~=~2--1) emission with $3\sigma$ or higher significance. The kinetic temperature $T_\mathrm{kin}$ and number density of molecular hydrogen $n$(H$_2$) of the post-shocked clouds located inside or on the shock front differed significantly from the other pre-shocked or ambient clouds; the cloud near the shock front shows the highest $T_\mathrm{kin}$ $\sim$50~K and the lowest $n$(H$_2$) $\sim$0.5--$0.7 \times10^3$~cm$^{-3}$, while the cloud $\sim$5~pc away from the edge of the SNR has the lowest $T_\mathrm{kin}$ $\sim$20~K and the densest $n$(H$_2$) $\sim$$1.3 \times10^3$~cm$^{-3}$. Since the spatial variation of $T_\mathrm{kin}$ cannot be explained by cosmic-ray heating alone or extra heating sources like O-type stars, we conclude that the high $T_\mathrm{kin}$ values of the clouds were caused by supernova shocks. This is the first compelling evidence of shock-heated molecular clouds in an extragalactic SNR.
\item The negative correlation between $T_\mathrm{kin}$ and $n$(H$_2$) indicates that clouds N, O, P, Q, and R have constant pressure and did not experience shock evaporation within the short SNR age of $\sim$4800~yr. On the other hand, clouds K, L, and T, which are on the inside or on the edge of the shell, are significantly shifted from this pressure-constant line to a higher-pressure region and have slightly reduced densities, suggesting that the shock-engulfed clouds were partially evaporated through the shock-cloud interaction. The decreased density and increased pressure (and kinetic temperature) of the shocked molecular clouds have important implications for understanding the negative feedback of energetic supernova shocks on star formation. Further follow-up studies are needed from both the observational and theoretical sides.
\end{enumerate}

\section*{Acknowledgements}
This paper makes use of the following ALMA data: ADS\/JAO.ALMA\#2015.1.01195.S and \#2021.2.00008.S. ALMA is a partnership of ESO (representing its member states), NSF (USA) and NINS (Japan), together with NRC (Canada), MOST and ASIAA (Taiwan), and KASI (Republic of Korea), in cooperation with the Republic of Chile. The Joint ALMA Observatory is operated by ESO, AUI/NRAO and NAOJ. The National Radio Astronomy Observatory is a facility of the National Science Foundation operated under cooperative agreement by Associated Universities, Inc. This paper employs a list of Chandra datasets, obtained by the Chandra X-ray Observatory, contained in~\dataset[DOI: 10.25574/cdc.161]{https://doi.org/10.25574/cdc.161. This research has made use of software provided by the Chandra X-Ray Center in the application package CIAO (v4.12).} This work was supported by JSPS KAKENHI grant No. 21H01136 (HS), 23H01211 (AB). This work was also supported by NAOJ ALMA Scientific Research Grant Code 2023-25A. This work was supported by the Ministry of Science, Technological Development and Innovations of Serbia, contract number is 451-03-47{/}2023-01{/}200002 (ML). PS acknowledges support via the Leiden University Oort Fellowship and the IAU - Gruber Foundation Fellowship. Support for CJL was provided by NASA through the NASA Hubble Fellowship grant No. HST-HF2-51535.001-A awarded by the Space Telescope Science Institute, which is operated by the Association of Universities for Research in Astronomy, Inc., for NASA, under contract NAS5-26555.

\software{IDL Astronomy User's Library \citep{1993ASPC...52..246L}, CASA \citep[v 5.4.1;][]{2022PASP..134k4501C}, MIRIAD \citep[][]{1995ASPC...77..433S}, CIAO \citep[v 4.12;][]{2006SPIE.6270E..1VF}, CALDB \citep[v 4.9.1;][]{2007ChNew..14...33G}, \texttt{MPFITEXY} \citep{2010MNRAS.409.1330W}.}

\facilities{Chandra, The Atacama Large Millimeter/Submillimeter Array (ALMA).}

\section*{APPENDIX\\Time evolution of the kinetic temperature}
To estimate the impact of the cosmic-ray heating on the kinetic temperature, 
we run one-dimensional plane parallel PDR models \citep{2022ApJ...926..148F}, which solve the time evolution of the gas temperature and abundances of chemical species self-consistently for given gas density distribution and dust-temperature distribution, considering gas-phase and gas-grain interaction (i.e., adsorption and desorption) and heating and cooling processes.
Elemental abundances are taken from \citet{2015ApJ...812..142A} (low metal abundances for the LMC in their Table 1). 
At the initial chemical state, we assume all hydrogen is in H${_2}$, all carbon is in CO, and the remaining oxygen is in H$_2$O ice.
The other elements are assumed to exist as either atoms or atomic ions in the gas phase.

The model simulation consists of two steps.
In the first step, the model simulates the temporal evolution of the gas temperature and chemical abundances for 10$^7$~yr, assuming the standard $\zeta$ value of $1\times10^{-17}$~s$^{-1}$.
The chosen timescale of 10$^7$~yr is arbitrary, but it is long enough for the C${^+}$, C$^0$, and CO abundances to reach steady-state values.
In the second step, the model follows the evolution for $1\times10^5$~yrs with an enhanced $\zeta$ value in a range between $3\times10^{-17}$~s$^{-1}$ and $1\times10^{-14}$~s$^{-1}$.
The first and second steps mimic conditions just before and after the supernova explosion, respectively.
As input parameters for the model, we employ the UV intensity $G_0$ = $110$ \citep{2012ApJ...744..160S} and the hydrogen (H{\sc i}+H$_2$) gas density $n_{\mathrm{HI+H_2}}$~=~$1\times10^3$~cm$^{-3}$.
The dust temperature is calculated as a function of $A_V$ using Equation 8 in \citet{2017A&A...604A..58H}.
The ratio of the hydrogen column density to $A_V$ is assumed to be $3.5\times10^{21}$~cm$^{-2}$ \citep{2006A&A...447..991C}.
The gas temperature evolution at the cloud center ($A_V = 5$ mag.) after the explosion is shown in Figure \ref{fig5}.

\bibliography{references}{}
\bibliographystyle{aasjournal}

\end{document}